\documentclass[twocolumn,prl,aps]{revtex4}
\usepackage{epsf}

\begin{document}

\title{ Comment on ``Absence of Compressible Edge Channel Rings in
Quantum Antidots'' }

\author{}

\address{}

\address{}

\date{\today}

\maketitle


\narrowtext

In a recent article, Karakurt {\em et al.}~\cite{KARAKURT}\
reported the absence of compressible regions~\cite{CHKLOVSKII}
around antidots in the quantum Hall regime. We wish to point out a
significant flaw in their analysis, which invalidates their claim.

The presence of compressible regions around antidots was proposed
by us~\cite{KATAOKA} in order to explain the so-called
``double-frequency Aharonov-Bohm oscillations''
\cite{FORD,SACHRAJDA}. The model considers Coulomb blockade
\cite{KATAOKA1} of tunnelling into compressible states formed
around an antidot. Karakurt {\em et al.}\ measured the temperature
dependence of the double-frequency Aharonov-Bohm resonances, and
fitted the data to two theories, one considering resonance through
a single state (Eq.~4 in Ref.~1) and one with multiple states
(Eq.~2) \cite{BEENAKKER}. The measured temperature dependence
matches that already observed in Ref.\ \cite{KATAOKA1} and follows
the behaviour predicted by the first theory, and they claim that
this shows that there are no compressible regions, in which
multiple states are pinned near the Fermi energy $E_{\rm{F}}$.

However, Karakurt {\em et al.}\ overlook the fact that the
multiple-state theory is only valid for a ladder of
single-particle states with a fairly constant density of states.
It predicts a temperature-independent tunnelling conductance
because thermal broadening increases the number of states involved
in tunnelling in proportion to the temperature $T$ whereas the
tunnelling through each state decreases as $1/T$.

\begin{figure}

\epsfxsize=8.5cm

\centerline{\epsffile{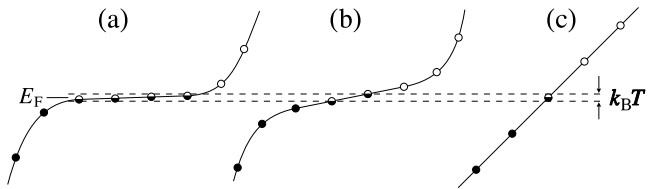}}

\caption{ The Landau level near the Fermi energy around an antidot
with (a) a well-defined compressible region around the antidot,
where all the compressible states are pinned within $k_{\rm{B}}T$;
(b) an incomplete compressible region, where its energy width is
larger than $k_{\rm B}T$, but the single-particle level spacing is
smaller than or comparable to $k_{\rm B}T$; (c) a steeply-sloping
potential, where a compressible region does not form and the level
spacing is much larger than $k_{\rm B}T$. }

\label{fig:adstates}

\end{figure}

It is not clear whether compressible regions should really exist
around the antidot, although their presence at high magnetic
fields is implied by our double-frequency model\ \cite{KATAOKA}.
Here, we consider two possible cases as depicted in
Fig.~\ref{fig:adstates}(a)~and~(b). The first case is with a
well-defined compressible region [Fig.~\ref{fig:adstates}(a)] and
fairly sharp transitions to incompressible regions (over a
distance of order the magnetic length). All the compressible
states stay within about $k_{\rm B}T$ of $E_{\rm{F}}$ \cite{LIER}.
Here, increasing $T$ does not change the number of states involved
in tunnelling, unless the increase is enough to involve
neighbouring incompressible states. Even so, as there are usually
many compressible states, involving a few more states would make
little difference, and hence a $1/T$ dependence is expected. The
results of Karakurt {\em et al.}\ cannot distinguish this
potential from a steeply-sloping potential, where the
single-particle level spacing is much greater than $k_{\rm B}T$
[Fig.~\ref{fig:adstates}(c)].

The second possibility is that the potential slopes more, since
screening
is imperfect [Fig.~\ref{fig:adstates}(b)]. Here, the
single-particle level spacing is smaller than or comparable to
$k_{\rm B}T$, but the energy width of the region of reduced slope
exceeds $k_{\rm B}T$. In this case, the multiple-state theory is
valid, and a temperature-independent tunnelling conductance is
expected. The results obtained by Karakurt {\em et al.}\ only
exclude such imperfect compressible regions.

We also wish to point out that we mentioned\ \cite{KATAOKA} that
our self-consistent model is only expected to work at relatively
large magnetic fields ($\sim 3$~T). It is very interesting to ask\
\cite{KATAOKA2} whether compressible regions should form fully at
the small fields ($<1$~T) used by Karakurt {\em et al.}.

In addition, the explanation given by Karakurt {\em et al.}\ for
double-frequency is pure speculation. They only {\em assume} that
there should be $i$ equally spaced resonances when $i$ Landau
levels form antidot states. While this would probably give $i$
resonances per $h/e$ of flux, there is no reason why they should
be equally spaced\ \cite{FORD,SACHRAJDA}. Also, this would give
$h/e$ oscillations when the constrictions were narrowed to filling
factor 1, whereas we observe the absence of any oscillations
(showing the $\nu=1$ plateau) until the constrictions are narrowed
enough for tunnelling via the lowest spin state (see Fig.~1 in
Ref.~3).

\vspace{10pt}

M. Kataoka and C. J. B. Ford

Cavendish Laboratory

Madingley Road,

Cambridge CB3 0HE, UK


\end{document}